\documentclass{aa}  
\usepackage[varg]{txfonts}
\usepackage{multirow}
\usepackage{graphicx}
\usepackage{mwe}
\usepackage{subcaption}
\usepackage{color}
\usepackage[colorlinks, citecolor={blue}]{hyperref}
\begin{document} 

\title{Understanding the role of morphology and environment on the dynamical evolution of isolated galaxy triplets}
\titlerunning{Understanding the role of morphology and environment on the  evolution of isolated galaxy triplets}

\author{
P.~V\'asquez-Bustos\inst{1}
\and 
M.Argudo-Fernandez\inst{1,2,3}
\and
D. Grajales-Medina\inst{4}
\and
S.~Duarte~Puertas\inst{5,6}
\and
S.~Verley\inst{2,3}
}

\institute{
Instituto de F\'isica, Pontificia Universidad Cat\'olica de Valpara\'iso, Casilla 4059, Valpara\'iso, Chile.  \\ \email{paulo.vasquez.b@mail.pucv.cl}
\and 
Departamento de F\'isica Te\'orica y del Cosmos Universidad de Granada, 18071 Granada, Spain.
\and
Instituto Universitario Carlos I de F\'isica Te\'orica y Computacional, Universidad de Granada, 18071 Granada, Spain
\and
Facultad de Ingenier\'ia, Departamento de Ciencias B\'asicas, Universidad Ean, Bogot\'a, Colombia
\and 
D\'epartement de Physique, de G\'enie Physique et d'Optique, Universit\'e Laval, and Centre de Recherche en Astrophysique du Qu\'ebec (CRAQ), Qu\'ebec, QC, G1V 0A6, Canada
\and
Instituto de Astrof\'{\i}sica de Andaluc\'{\i}a - CSIC, Glorieta de la Astronom\'{\i}a s.n., 18008 Granada, Spain
}

\date{Received 26 October 2022 / Accepted 17 November 2022}

 
 \abstract
  {The environment where galaxies reside affects their evolutionary histories. Galaxy triplets (systems composed of three physically bound galaxies) constitute one of simplest group of galaxies and are therefore excellent laboratories to study evolutionary mechanisms where effects of the environment are minimal.} 
  {We present a statistical study of the dynamical properties of isolated galaxy triplets as a function of their local and large scale environments. To explore the connection of the dynamical evolution on the systems with the evolution of the galaxies composing the triplets, we consider observational properties as morphology and star formation rate (SFR).}
  {We used the SDSS-based catalog of Isolated Triplets (SIT), which contains 315 triplets. We classified each triplet according to galaxy morphologies and defined a parameter $Q_{\rm trip}$ to quantify the total local tidal strengths in the systems. To quantify the dynamical stage of the system we used the parameters harmonic radius, $R_H$, velocity dispersion, $\sigma_{vr}$, crossing time, $H_0t_c$, and virial mass, $\rm M_{vir}$.}
   {Triplets composed of three early type galaxies present smallest $R_H$, indicating that they are in general more compact than triplets with one or more late type galaxies. Among triplets with low values of $R_H$ and $H_0t_c$, SIT triplets with $Q_{\rm trip}$~<~$-$2 are relaxed systems, more dynamically evolved, while triplets with $Q_{\rm trip}$~>~$-$2 show compact configurations due to interactions within the system, such as on-going mergers.}
   {We found that there is no dominant galaxy in triplets in terms of properties of stellar populations such as global colour and SFR. Moreover, the global SFR in isolated triplets composed of two or more early-type galaxies increases with the stellar mass ratio of the galaxies with respect to the central galaxy, therefore the system is globally ’rejuvenated’.}
   
\maketitle
%

\section{Introduction}

The distribution of galaxies in the universe is not homogeneous, it has been found that galaxies tend to group in structures composed of N galaxies where the most common are clusters with hundreds of galaxies \citep{1994AJ....107.1623R, 1999ApJ...527...54B,2012A&A...545A.104L}. One of the simplest groups are galaxy triplets \citep{1979AISAO..11....3K,2000ASPC..209...11K,2009MNRAS.394.1409E,2009AstBu..64...24M}, i.e. systems composed of three physically bounded galaxies (N~=~3). Originally, galaxy triplets were considered to be systems composed of a galaxy pair with a remote galaxy, however, recent studies have found that the majority of galaxy triplets show evidence of a long dynamical evolution, where the system is embedded in a common dark matter halo, and their member galaxies present similar properties \citep{2000MNRAS.319..851C,2011AJ....141...74H,2013MNRAS.433.3547D,2016RAA....16...72F,2016ARep...60..397E}.

Indeed, recent analyses suggest that galaxy triplets composed of three luminous galaxies, should not be considered as analogous of galaxy pairs with a third extra member, but rather as a natural extension of compact groups \citep{2015MNRAS.447.1399D,2016MNRAS.459.2539C}. A compact group (CG) is a dense isolated galaxy system, composed of at least four galaxies, separated by a projected distance on the order of their size (50-100\,kpc), and mean velocity dispersion about 230\,km\,s$^{-1}$ \citep{1982ApJ...255..382H,1994AJ....107.1235P,2009MNRAS.395..255M,2020ApJS..246...12Z}. As in CGs, the high compactness in galaxy triplets might promote strong interactions between the galaxy members, making galaxy triplets ideal laboratories for studying the environmental effects on galaxy evolution \citep{2018MNRAS.481.2458D,2021MNRAS.504.4389D}.

The are a few catalogues of galaxy triplets in the bibliography, being one of most studied the catalogues compiled by \cite{1979AISAO..11....3K,1995AZh....72..308T,2012MNRAS.421.1897O}, henceforth K-triplets, W-triplets (from wide triplets), and O-triplets, respectively; and the SDSS catalogue of isolated triplets (the SIT catalogue) compiled by \citet{2015A&A...578A.110A}. Using the K-triplets, W-triplets, and O-triplets
\citet{2013MNRAS.433.3547D} found that galaxies in triplets have similar properties (star formation rate, colors, and stellar population) to galaxies in compact groups. In particular, \citet{2015MNRAS.447.1399D} suggest that the formation of elliptical galaxies through mergers in galaxy triplets could be favoured by the configuration and dynamics of long term evolved systems. In general galaxy triplets with similar photometric properties reside and evolve in a common dark matter halo. In addition, \citet{2019MNRAS.482.2627T} found that the dynamics of isolated triplets is slightly different that in K-triplets. Therefore, it is still not well understood how the dynamics of these systems affect the evolution of the galaxies, since it might be influenced by their large scale environment.

Galaxy triplets can be found both within larger structures, as large groups and clusters, and in the field. In fact, \citet{2011MNRAS.418.1409M} found that about half of CGs are embedded sub-structures of rich groups or clusters, while the other half are isolated structures or associated with comparably poor groups. Therefore, in order to better understand the evolution of the galaxies in triplets, it would be appropriate to consider isolated triplets (i.e., systems without close neighbours that might appreciably exert any influence during a past crossing time $\rm t_{cc}\,\approx\,$3\,Gyr) to minimise the possible effects of neighbour galaxies in their large-scale environment. In this context, \citet{2015A&A...578A.110A} claimed that there is no difference in the degree of interactions with the large-scale environment for isolated galaxies, isolated pairs and the SIT, which suggests that they may have a common origin in their formation and evolution. Therefore any difference in their observed properties is due to the influence of their local environment and the dynamic of the systems.

In this work we perform a statistical study of the dynamical properties of isolated galaxy triplets in the SIT as a function of their local environment. To explore the connection of the dynamical evolution of the systems with the evolution of the galaxies composing the triplets, we consider observational properties such as morphology and star formation rate (SFR). This work is organised as follows. In Sect.~\ref{Sec:data} we describe the SIT and how we classify the systems according to morphology and SFR. We also present the parameters we use to quantify the dynamical stage of the triplets, and to quantify both, the local and large-scale structure (LSS) environment, as well as a new defined parameter to quantify the total local tidal strength of the triplet system. In Sect.~\ref{Sec:Res} we present our results in terms of environments, dynamics, and their interconnections with galaxy properties; with the corresponding discussion in Sect.~\ref{Sec:dis}. Finally, we present a summary of the work and our conclusions in Sect.~\ref{Sec:con}. Throughout the study, a cosmology with $\Omega_{\Lambda_{0}} = 0.7$, $\Omega_{\rm m_{0}} = 0.3$, and $H_0=70$\,km\,s$^{-1}$\,Mpc$^{-1}$ is assumed.

\section{Data and methodology} 
\label{Sec:data}

\subsection{The SIT}

\begin{figure}
    \centering
    \includegraphics[width=\columnwidth]{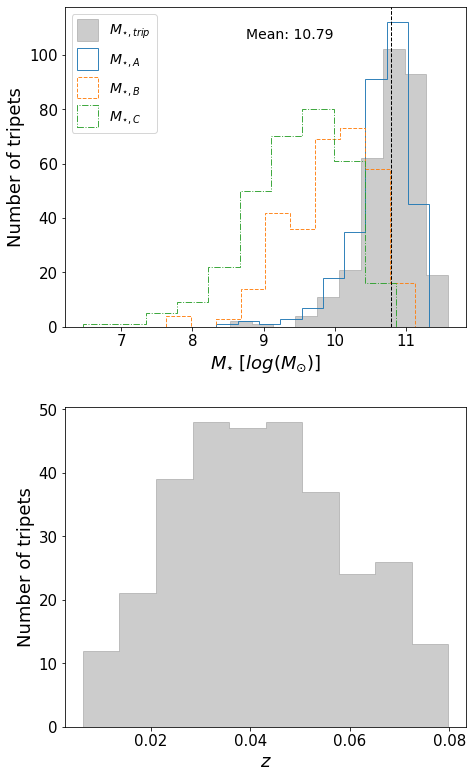}
    \caption{Properties of the galaxies in the SIT. \textit{Upper panel:} Distribution of stellar masses of the A ($M_{\star,A}$), B ($M_{\star,B}$), and C galaxies ($M_{\star,C}$), in different colour and line-styles according to the legend, with respect to the total stellar mass of the isolated triplet ($M_{\star, trip}~=~M_{\star,A}~+~M_{\star, B}~+~M_{\star, C}$), as filled grey histogram. The vertical black dashed line indicates the mean value of the $M_{\star, trip}~=~10.79~\log(M_{\odot})$. \textit{Lower panel:} Redshift distribution of the SIT.}
    \label{fig:sit}
\end{figure}

As introduced in the previous section, we used for our study the SIT sample of isolated galaxy triplets \citep{2015A&A...578A.110A}. The sample is based on photometric and spectroscopic data from the Tenth public Data Release of the Sloan Digital Sky Survey \citep[SDSS-DR10;][]{2014ApJS..211...17A}, for galaxies with r-band model magnitudes within $14.5~\leq~m_{r}~\leq~17.7$ and redshift $0.005~\leq~z~\leq~0.080$ (see the lower panel in the Fig.~\ref{fig:sit}). 

The galaxies in the SIT are classified from the brightest to the faintest galaxy, refereed as the A, B, and C galaxy respectively. Under this definition, the A galaxy is expected to be the most massive galaxy in the triplets (as shown in the upper panel in the Fig.~\ref{fig:sit}). SIT triplets usually show a hierarchical structure, as shown in Fig.~\ref{fig:sit}, where the mass of the B galaxy is typically one-tenth of the mass of the A galaxy, and the mass ratio of the A and C galaxies is about 1/100. 

The triplets in the SIT are selected under a restricted isolation criterion, where the system has no neighbours up to a projected physical distance of 1\,Mpc within a line-of-sight velocity difference $\Delta\,v~\leq~500$~km~s$^{-1}$. The B and C galaxies satisfy $\Delta\,v~\leq~160$~km~s$^{-1}$ within a projected distance $d~\leq~450$\,kpc from the A galaxy. The SIT on \citet{2015A&A...578A.110A} is composed of 315 isolated triplets, corresponding to 3\% of the local Universe \citep[$z~\leq~0.080$ ][]{2015A&A...578A.110A}. 

\subsection{Morphology}
\label{Sec:morf}

Since the SIT is based on SDSS data, we use the morphology classification from \citet{2018MNRAS.476.3661D}, which provides deep learning based morphological classification for $\sim$670\,000 galaxies from the SDSS DR7 \citep{2009ApJS..182..543A}. The Deep Learning models used in \citet{2018MNRAS.476.3661D} were trained and tested using visual morphological classification from the \textit{Galaxy Zoo} project\footnote{\texttt{\url{www.galaxyzoo.org}}}, in particular from the Galaxy Zoo 2 \citep[GZ2,][]{2013MNRAS.435.2835W} catalogue; and the morphological classification by \citet{2010ApJS..186..427N}.
\citet{2018MNRAS.476.3661D} provides a T-type classification that ranges from -3 to 10, where T-type~$\leq$~0 corresponds to early-type galaxies (i.e., elliptical and lenticular galaxies), and positive values to late-type galaxies, with T-type~=~10 for irregular galaxies. 

We find that all galaxies in the SIT have morphology classification in \citet{2018MNRAS.476.3661D}. For the purpose of this work, we only consider if the galaxies are classified as late-type (including irregulars) or early-type (considering ellipticals and lenticulars). We find 359 galaxies classified as early-type (38\%), and 586 as late-type (62\%). 
We therefore classified triplets in the SIT in four categories according to the morphology of the galaxies as follows:

\begin{itemize}
    \item TL: the three galaxies in the triplet are late-
type galaxies, 65 triplets (20.6\% of triplets in the SIT). 
    \item TE: the three galaxies in the triplet are early-type galaxies, 28 triplets (8.9\% of triplets in the SIT). 
    \item TCL: triplets with the central galaxy a late-type, 70 triplets (22.2\% of triplets in the SIT). 
    \item TCE: triplets with the central galaxy an early-type, 148 triplets (47.0\% of triplets in the SIT). 
\end{itemize}

\begin{figure*}
    \centering
    \includegraphics[width=.45\textwidth]{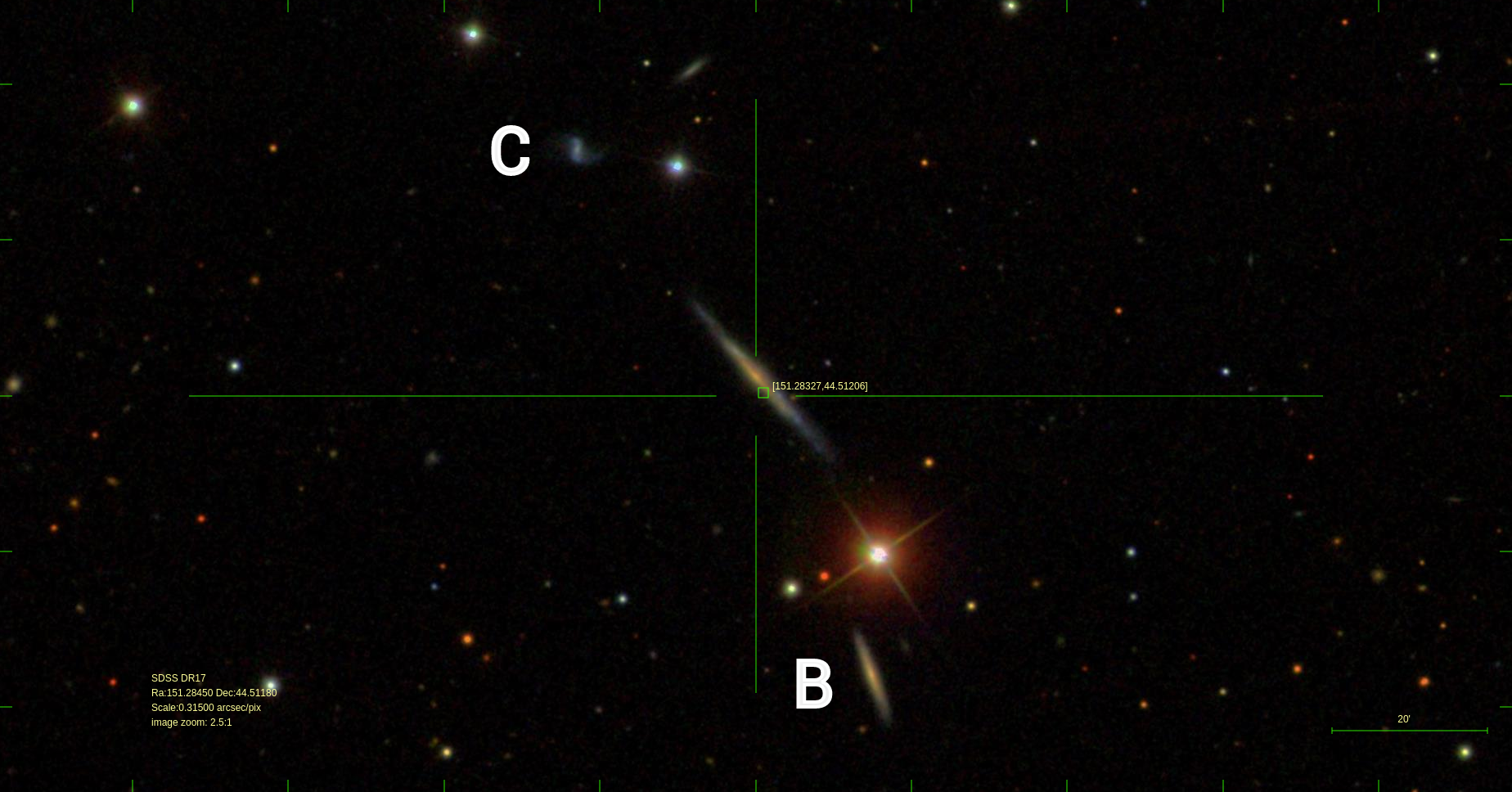}
    \includegraphics[width=.45\textwidth]{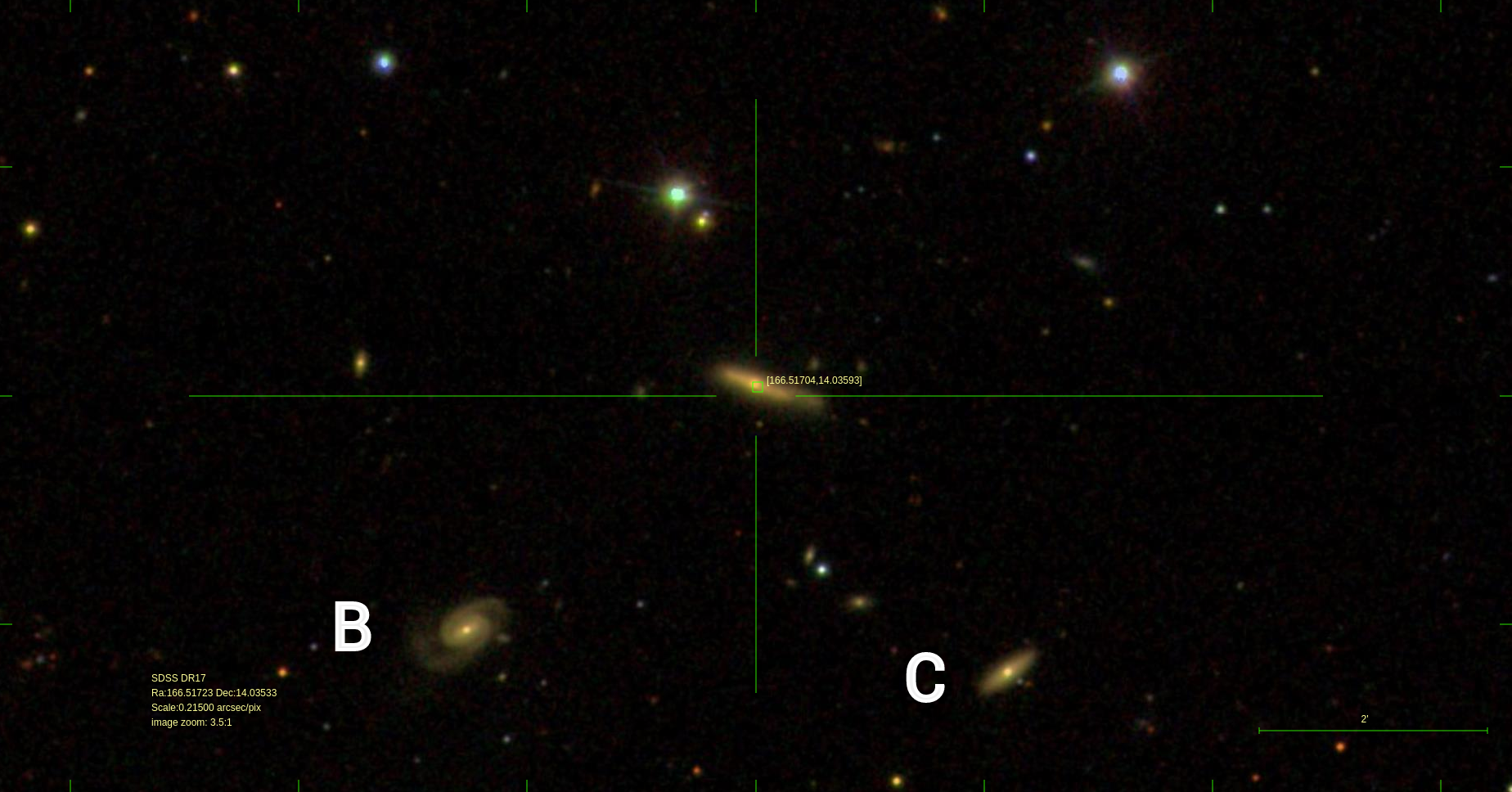}
    \includegraphics[width=.45\textwidth]{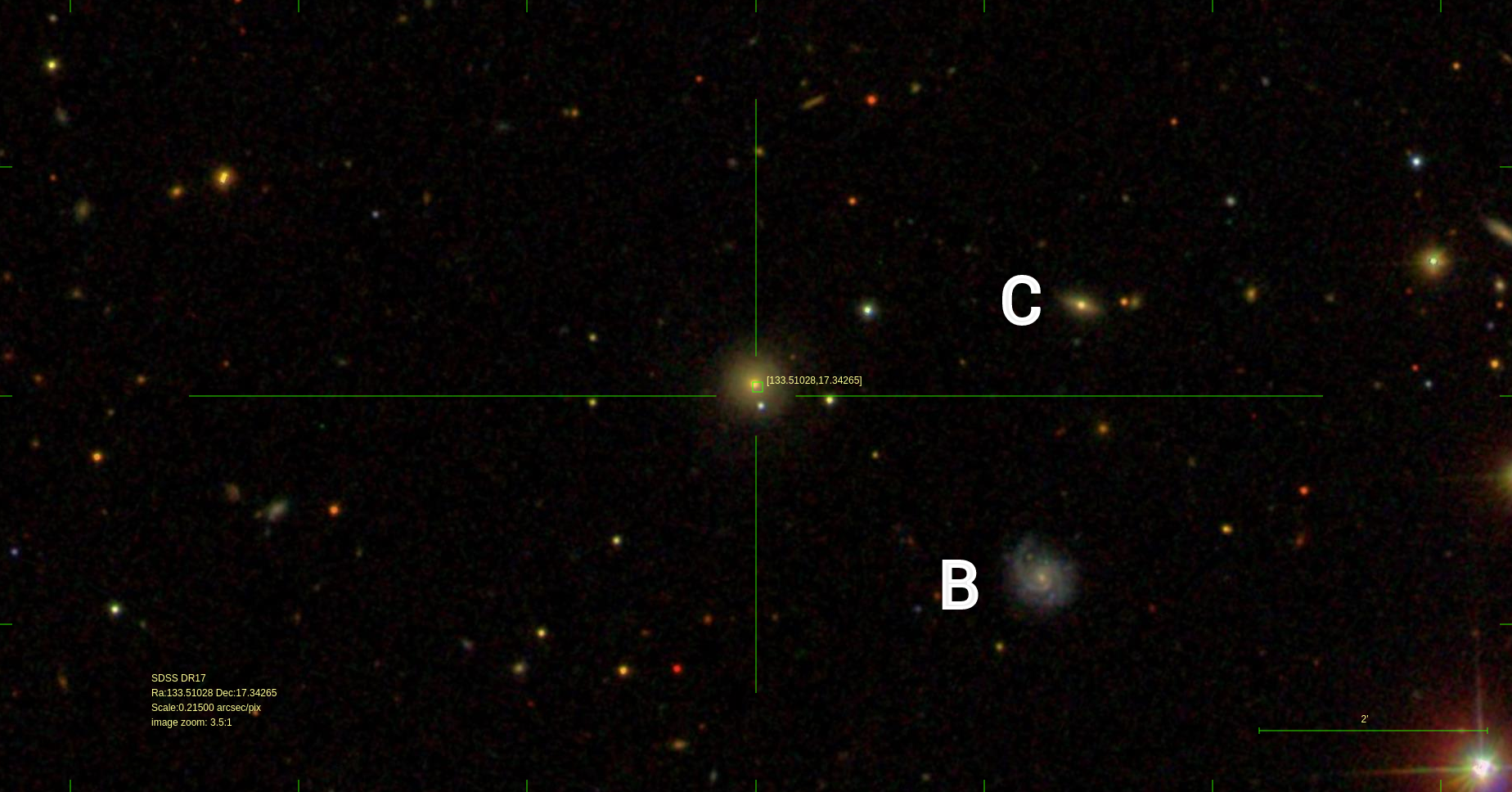}
    \includegraphics[width=.45\textwidth]{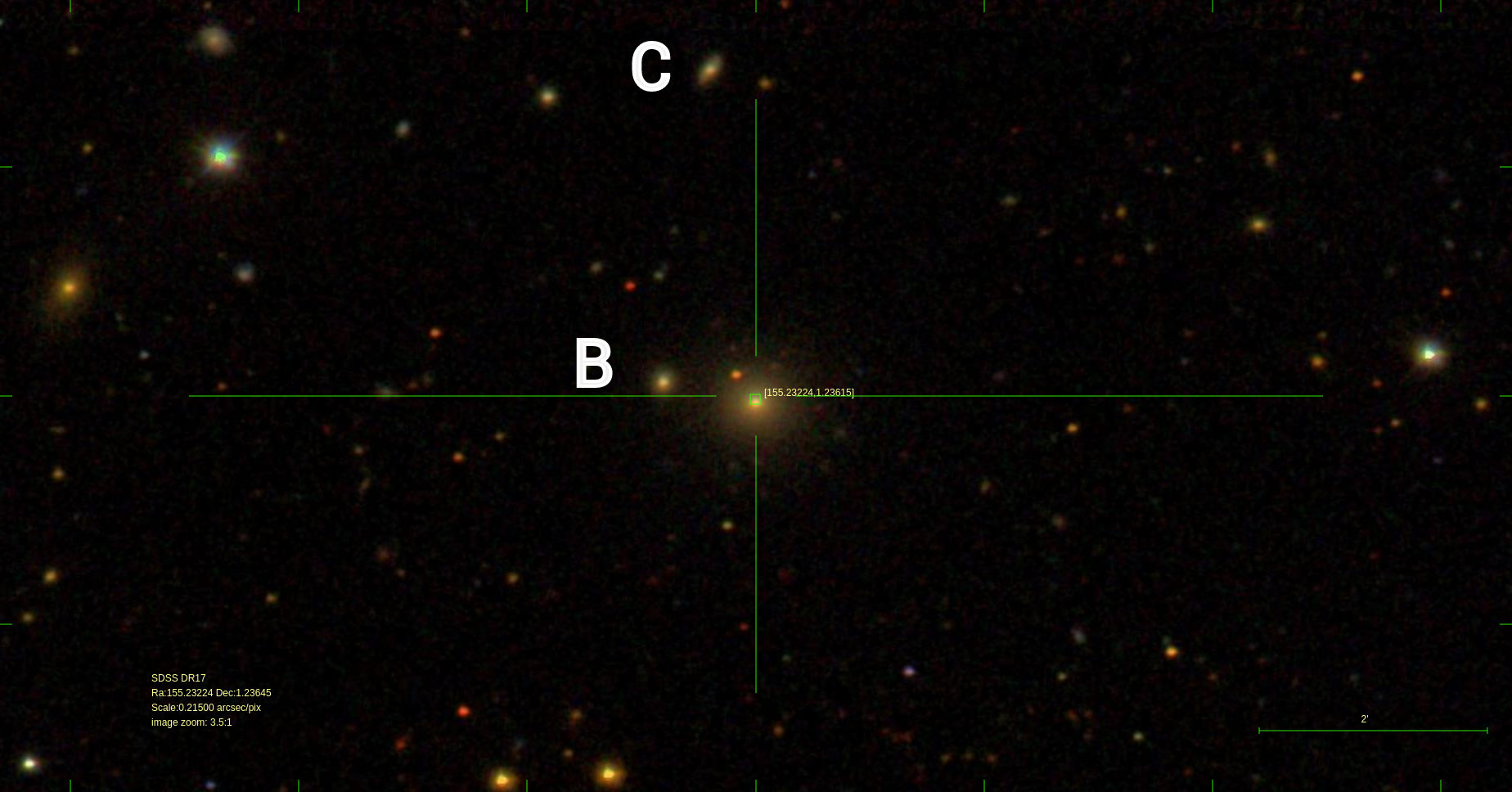}
    \caption{SDSS-DR17 three colour images of four selected SIT systems (SIT 86, SIT 222, SIT 217, and SIT 17 from the upper left to lower right panels) as an example of the morphological categories classification, when: a) the system is composed of three spiral galaxies (TL, upper left panel); b) the system is composed of three elliptical galaxies (TE, upper right panel); c) the A galaxy is a spiral galaxy (TCL, lower left panel); and d) the A galaxy is an elliptical galaxy (TCE, lower right panel). North is up and east is left. Image scale 0.215\arcsec/pixel and the center on the image represent the A galaxy}\label{fig:sit-examples}
\end{figure*}

The TCL and TCE categories are based on the morphology on the A galaxy, when the three galaxies do not have the same morphology. An example of SIT triplets classified in each morphology category is shown in Fig.~\ref{fig:sit-examples}.

\subsection{Star formation rate}
\label{Sec:sfr}

For the purpose of this work, we use the GALEX-SDSS-WISE Legacy Catalog \citep[GSWLC,][]{2016ApJS..227....2S}, which contains physical properties, like the star formation rate (SFR), of  $\sim 700,000$ galaxies with SDSS redshift below 0.30 (0.01\,$<$\,z\,$<$\,0.30) and magnitude $m_r\,<$\,18\,mag. The GSWLC contains galaxies within the Galaxy Evolution Explorer (GALEX) satellite All Sky Survey \citep{2005ApJ...619L...1M} footprint, regardless of a  detection, altogether covering 90\% of the SDSS footprint. However there are three versions of the catalog (GSWLC-A, M, D), depending on the depth of the ultraviolet (UV) photometry. 

\citet{2016ApJS..227....2S} used multi-wavelength observations from the UV to the infrared (IR) to construct the spectral energy distribution (SED) and derive the physical properties (SFRs, stellar masses, and dust attenuation) with the \textit{CIGALE} code \citep{2005MNRAS.360.1413B,2009A&A...507.1793N,2019A&A...622A.103B}, using state-of-the-art UV/optical SED fitting techniques. In particular we use the second version of the catalogue, GSWLC-2 \citep{2018ApJ...859...11S}, where the mid-infrared flux from 22 micron (or 12 micron, if 22 micron is not detected) observations with the Wide-field Infrared Survey Explorer \citep[WISE,][]{2010AJ....140.1868W} are used in the SED fitting jointly with UV/optical photometry to derive more accurate SFRs. 
We used the Medium-deep catalogue, GSWLC-M2 that covers 49\% of SDSS dataset. The numbers of SIT triplets with SFR information for the three galaxies in each morphological category are:

\begin{itemize}
    \item TL: 38 triplets (58.5\% of SIT triplets classified as TL).
    \item TE: 13 triplets (46.4\% of SIT triplets classified as TE).
    \item TCL: 33 triplets (47.1\% of SIT triplets classified as TCL).
    \item TCE: 58 triplets (39.2\% of SIT triplets classified as TCE).
\end{itemize}

\subsection{Quantification of the environment}
\label{Sec:env}

\citet{2015A&A...578A.110A} quantified the LSS environment and the level of interaction between the galaxies in the system, based on two complementary parameters: the projected number density ($\rm \eta_{k,LSS}$), and  tidal strength ($\rm Q$) local and on LSS.

\begin{itemize}
    \item Projected number density:
    This parameters is the density of neighboring galaxies around the A galaxy defined by the distance to the 5$^{th}$ nearest neighbour \citep{2007A&A...472..121V}. 
    
    \begin{equation}
     \eta_{k,LSS}=\log\left(\frac{k-1}{V(d_k)}\right) \quad,
    \end{equation}
    
    with $V(d_k)=\frac{4}{3}\pi d_k^3$ where $d_k$ is the projected distance to $k^{th}$  nearest neighbour, with $k\leqslant5$ this depending of the neighbors in the field out to a projected distance of 5\,Mpc.
    
    \item Tidal strength: The parameter $Q_A$ is an estimation of the sum of all the gravitational tidal force with respect to the central galaxy \citep{2007A&A...472..121V,2013MNRAS.430..638S,2015A&A...578A.110A}.
    
    \begin{equation}\label{Eq:Q} 
    Q_A \propto \log\left[\sum_{i}\frac{M_{\star i}}{M_{\star A}}\left(\frac{D_{A}}{d_{i}}\right)^{3}\right] \quad,
    \end{equation} 
    
    where $M_{\star A}$ and $D_{A}$ are the stellar mass and the diameter of the A galaxy, respectively, $M_{\star i}$ is the stellar mass of the i$^{th}$ neighbour galaxy, and $d_{i}$ is its projected distance to the A galaxy.
\end{itemize}

\subsection{Dynamical parameters}
\label{Sec:dyn}

We used the following parameters to quantify the dynamic configuration of the triplets in the SIT, as defined in \citet{1982ApJ...255..382H,1992ApJ...399..353H}, which provide an estimation of the evolution stage on the systems.  

\begin{itemize}
\item Harmonic radius ($\rm R_H$): This parameter represents a measurement of the effective radius of a galaxy group or cluster. 
\begin{equation}
    R_H = \left( \frac{1}{N}\sum R_{ij}^{-1}\right)^{-1},
\end{equation}
where $R_{ij}$ are the projected galaxy-galaxy separations, in Mpc.

\item Velocity dispersion ($\rm \sigma_{v_r}^2$): This parameter corresponds to the dispersion of the radial velocities of the galaxies in the triplet:
\begin{equation}
    \sigma_{v_r}^2 = \frac{1}{N-1}\sum \left(v_r-\left \langle  v_r\right \rangle \right)^2\;,
\end{equation}
with $v_r \propto c z$.

\item Crossing time ($\rm H_0t_c $): This parameters is a measure of the time that takes a single galaxy to across their system. We expressed the crossing time with respect to another dynamical parameters like harmonic radius and the velocity dispersion, this expression is normalised by cosmological terms.
\begin{equation}
    H_0t_c = \frac{H_0\,\pi \,R_H}{\sqrt{3}\,\sigma_{v_r}}.
\end{equation}

\item Virial Mass ($\rm  M_{vir} $): The mass estimation of a gravitationally bound system of triplets is related to the radius of the system and the velocity dispersion, and a factor to account for the number of galaxies in the system. This estimator represents an approximation of the halo mass and barionic mass.

\begin{equation}
    M_{vir} = \frac{3\pi\, N\, R_H\, \sigma^2_{v_r}}{\left(N-1\right)G}.
\end{equation}
\end{itemize}

The result of the quantification of the dynamical parameters for the SIT is presented in Sect.~\ref{Sec:res-dynparams}.

\section{Results}
\label{Sec:Res}

\subsection{Environment of galaxy triplets}
\label{Sec:res-environment}

The tidal strength Q (Eq.~\ref{Eq:Q}) parameter provided by \citet{2015A&A...578A.110A} estimates the total tidal strengths exerted on the A galaxy in the SIT triplets. 

\begin{figure}
    \centering
    \includegraphics[width=\columnwidth]{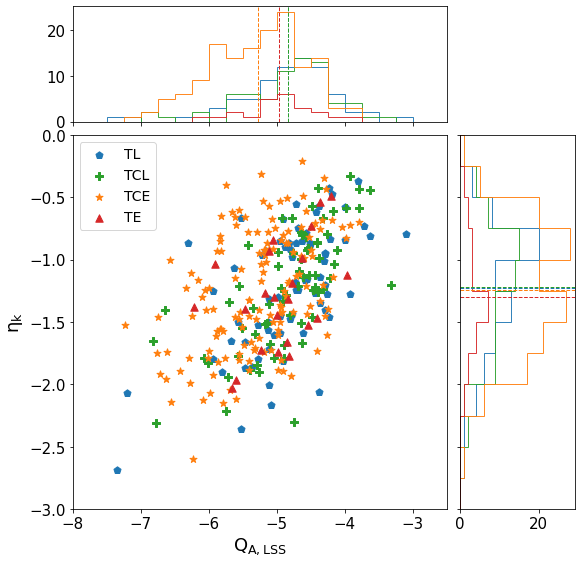}
    \caption{ Quantification of the LSS environment for the SIT. The morphological category for each triplet is depicted with different colours and markers. Blue pentagons correspond to triplets where the three galaxies are spirals (TL); green crosses correspond to triplets where the A galaxy is spiral (TCL); orange stars correspond to triplets where the A galaxy is elliptical (TCE); and red triangles for triplets where the three galaxies on are ellipticals (TE), as indicated in the legend. The distribution and their mean values of the parameters $Q_{\rm A,LSS}$ and $\eta_{k,LSS}$ for each morphological category is shown in the upper and right panels respectively, following the same colour code.}
    \label{fig:LSS}
\end{figure}

To quantify the tidal strength due to the LSS environment, $Q_{\rm A,LSS}$ considers all neighbour galaxies within $\Delta\,v~\leq~500$~km~s$^{-1}$ and up to a projected distance of 5\,Mpc (without considering the companion galaxies in the triplet). Figure~\ref{fig:LSS} shows the diagram $\eta_{k,LSS}$ versus $Q_{\rm A,LSS}$, similar to the lower right panel of Fig.\,3 in \citet{2015A&A...578A.110A}, but for SIT triplets only, colour coded according to the morphological classification of the triplet. Note that these parameters do not take into account the influence of the B and C galaxies, therefore they quantify the large-scale environment of the system. SIT triplets with lower values of both parameters are more isolated from their LSS environment. In general, we do not find a relation with the morphology of the triplets. However, as shown in the distribution of the $Q_{\rm LSS}$ in Fig.~\ref{fig:LSS}, TCE triplets for a fixed $\eta_{k}$ present lower average $Q_{\rm LSS}$ values and large scatter than the other categories.

\begin{figure}
    \centering
    \includegraphics[width=\columnwidth]{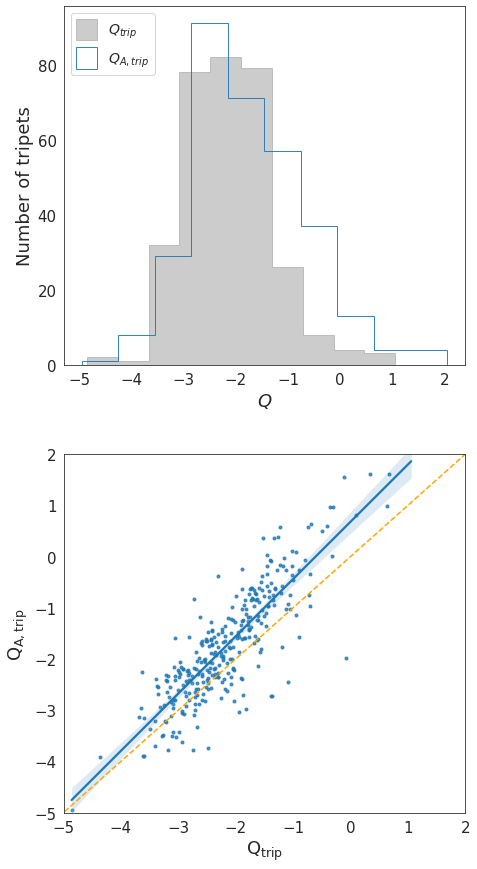}
    \caption{Comparison of the new total local tidal strength parameter $Q_{trip}$ with respect to the classic $Q_{A,trip}$ parameter in \citet{2015A&A...578A.110A} for SIT isolated triplets. \textit{Upper panel}: Distribution of the $Q_{trip}$ parameter (filled grey histogram) in comparison to the distribution of the $Q_{A,trip}$ parameter (blue line histogram). \textit{Lower panel}: Scatter plot of the $Q_{trip}$ parameter with respect to the $Q_{A,trip}$ parameter (blue circles). The orange dashed line represents the y=x relation, the blue line represents the data linear regression, and the shaded light blue area indicates its corresponding error.}
    \label{fig:Qtrip}
\end{figure}

On the other hand, $Q_{A,trip}$ is the local tidal strength considering the influence of the B and C galaxies on the A galaxy. This definition allows \citet{2015A&A...578A.110A} to compare the tidal parameters on central galaxy among isolated galaxies, isolated pairs, and isolated triplets, however it does not necessarily account to the total local tidal strengths in the triplet (i.e., considering the tidal strengths on each galaxy of the triplet). In this regard, we defined a new parameter, $Q_{trip}$, which considers the total local tidal strengths as the mean of the local tidal strengths of each galaxy in the triplets as follow:
\begin{equation}\label{Eq:Qtrip}
    Q_{trip} = \frac{Q_{A,trip} + Q_{B,trip} + Q_{C,trip}}{3} \quad,
\end{equation}

with $Q_{B,trip}$ being the local tidal strength considering the influence on A and C galaxies on the B galaxy, and $Q_{C,trip}$ considering the influences on A and B galaxies on the C galaxy, respectively. The values of the estimation of the $Q_{trip}$ parameter for each SIT triplet can be found in Table~\ref{tab:table}. The comparison of the new defined $Q_{trip}$ parameter with respect to the classic $Q_{A,trip}$ is shown in Fig~\ref{fig:Qtrip}. Note that there is no need to define a new $Q_{\rm A,LSS}$ since the closest LSS associations are located at projected distances $d~\geq~1$\,Mpc from the triplets, therefore an average $Q_{\rm LSS}$ would provide same values at significant level.

\begin{table*}[]
    \centering
    \begin{tabular}{ccccccccc}
        \hline\hline
        (1)&(2)&(3)&(4)&(5)&(6)&(7)&(8)&(9)\\
         SIT&RA&DEC&$Q_{trip}$&$R_H$&$\sigma_{v_r}$&$H_0t_c$&log$(M_{vir})$&Category \\
         &(deg)&(deg)&&(kpc)&(kms$^{-1}$)&&(log($M_{\odot}$))&\\
         \hline
         \\
         1 &	171.3154 &	-2.0761 &	-2.93 &	374.92 &	20.88 &	2.28 &	11.73 &	TCE\\
         2 &	208.6770 	&65.2443 &	-2.09 &	184.60 &	77.38 &	0.30 &	12.56 &	TCE\\
         3 	&211.2034& 	4.9810 &	-1.79 	&98.58 &	81.46 &	0.15 &	12.33 &	TCL\\
         4 	&251.4674 &	44.4374 &	-2.77 &	342.22 &	55.90 &	0.78 &	12.56 &	TCL\\
         5 &	133.5354 	&0.4985 &	-2.75 &	203.72 &	24.65 &	1.05 &	11.61 &	TCL \\
         6 &	207.8236 &	0.3849 &	-1.08 &	54.13 &	34.49 &	0.20 &	11.33 &	TCE \\
         7 &	218.9557 	&62.4211 	&-2.70 &	395.53& 	75.03 &	0.67 &	12.86 &	TL\\
         8 &	197.6862 &	0.0320 &	-3.24 &	373.01 &	40.84 &	1.16 &	12.31 &	TL\\
        9 &	320.8764 	&-7.7459 &	-2.29 &	46.50 &	60.43 &	0.10 &	11.75 &	TCE\\
        10& 	22.2240 &	14.7207 &	-1.27 &	124.35 	&102.74 	&0.15 &	12.63 &	TCL
        \\
        ...&...&...&...&...&...&...&...&...\\
         \hline
    \end{tabular}
    \caption[Dynamical parameters and morphology of SIT isolated triplets]{Dynamical parameters and morphology of SIT isolated triplets. The full table is available in electronic form at the CDS. The columns correspond to: (1) isolated triplet identification; (2) J2000.0 right ascension in degrees; (3) J2000.0 declination in degrees; (4) $Q_{trip}$, tidal strength estimation of the triplet due to member galaxies; (5) $R_H$, harmonic radius in kpc ; (6) $\sigma_{v_r}$, velocity dispersion in kms$^{-1}$; (7) $H_0t_c$, crossing time; (8) log($M_{vir}$), virial mass in log($M_{\odot}$); and (9) morphology classification of the isolated triplet (0: TL; 1: TCL; 2: TCE; 3: TE).}
    \label{tab:table}
\end{table*}

\subsection{Dynamics of galaxy triplets}
\label{Sec:res-dynparams}

We have estimated the dynamical parameters harmonic radius ($R_H$), velocity dispersion ($\sigma_{v_r}$), crossing time ($H_0 t_c$), and virial mass ($M_{vir}$) for the SIT, as defined in Sect.~\ref{Sec:dyn}. The values of the parameters for each SIT triplet can be found in Table~\ref{tab:table}. 

\begin{figure*}
    \centering
        \includegraphics[width=.45\textwidth]{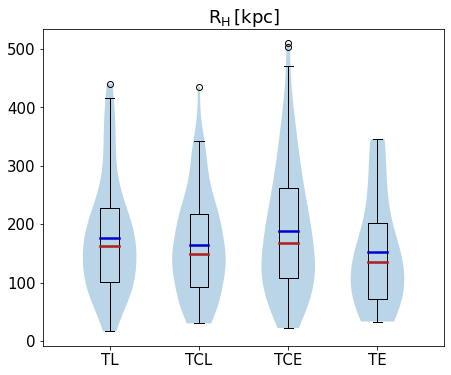}
        \includegraphics[width=.45\textwidth]{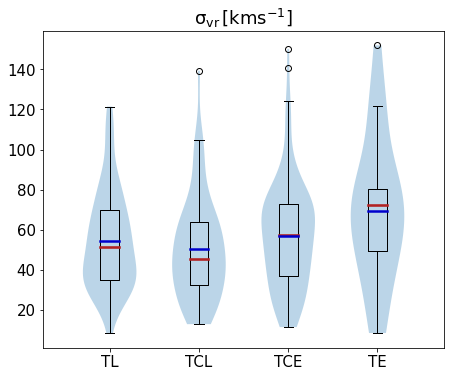} \\
        \includegraphics[width=.45\textwidth]{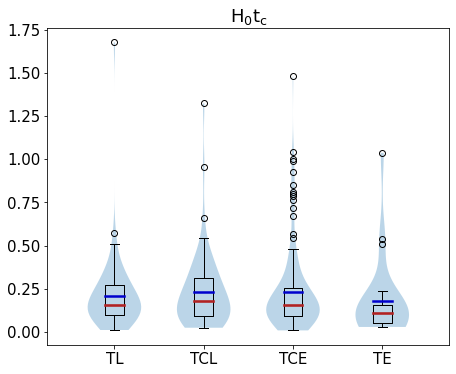}
        \includegraphics[width=.45\textwidth]{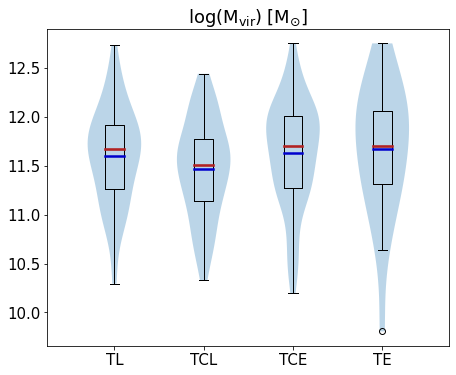}
        \caption{Box plots of the dynamical parameters for the SIT, from upper left to lower right: harmonic radius ($R_H$), velocity dispersion ($\sigma_{v_r}$), crossing time ($H_0 t_c$), and virial mass ($M_{vir}$). According to the dynamical parameters, the pale light blue shape represent to density diagram (violin plot), that show for each dynamical parameter the form of data distribution. The inner box on the violin plot is a representation of interquartile range (IQR) of the median (red horizontal line) and its 95$\%$ of confidence intervals. We also show the mean values (blue horizontal line) and the outliers points of the distributions as black open circles, which represent the atypical points of the samples.}
        \label{fig:violin}
\end{figure*}

The violin-box plots in Fig.~\ref{fig:violin} provide a global view of the quantification of the dynamical parameters as a function of the morphology of the SIT isolated triplets. In addition, the median values of the parameters for each morphological category, as defined in Sect.~\ref{Sec:morf}, are shown in Table~\ref{tab:dyn_tab}. For comparison, the median value of the dynamical parameters for the SIT are $R_H$~=~157.59\,kpc, $\sigma_{v_r}$~=~53.90\,kms$^{-1}$, $H_0t_c$~=~0.38, and $log(M_{vir})$~=~12.13\,$M_{\odot}$, within an interquartile range $R_H$~=~131.60\,kpc, $\sigma_{v_r}$~=~36.12 kms$^{-1}$, $H_0t_c$~=~0.43, and $log(M_{vir})$~=~0.73 $M_{\odot}$. 

We found that TE triplets have the smallest $R_H$ value (153.22\,kpc), as expected according to \citet{1982ApJ...255..382H}, which indicates that isolated triplets composed of three elliptical galaxies are in general more compact than triplets with one or more spiral galaxies. We also found that TL and TCL triplets in general present lower values of the velocity dispersion than TCE and TE triplets. We discuss our results in more detail in Sect.~\ref{Sec:dis-dynparam}.

\begin{table}
    \centering
    \setlength{\arrayrulewidth}{0.1mm}
    \begin{tabular}{ccccc}
    \hline\hline
       & (1) & (2) & (3) & (4) \\
     Category & $R_{\rm H}$ & $\sigma_v$  & $H_0t_c$  & $\rm log(M_{\rm vir})$  \\ 
      & (kpc) & (km\,s$^{-1}$)  &  & (log\,$M_{\odot}$) \\ 
    \hline
    Table A&&&&\\
    TL&177.34&54.44&0.21&11.60\\
    TCL&164.05&50.15& 0.23&11.48\\
    TCE&188.03&56.97& 0.23&11.63\\
    TE&153.22&69.27&0.18&11.68\\
    \hline
    Table B&&&&\\
    TL& 127.13&34.81&0.17&0.66\\ 
    TCL& 124.10&31.12&0.22&0.63\\
    TCE&152.99&35.61&0.16&0.74\\
    TE&130.68&30.95&0.10&0.75\\
     \hline
     \end{tabular}
    \caption[Statistics of the dynamical parameters of the SIT by morphology]{Statistics of the dynamical parameters of the SIT by morphology. Median values (Table A) and uncertainties (given by the interquartile range, Table B) of the dynamical parameters of the SIT, for each morphological type. The columns correspond to: (1) harmonic radius, in kpc; (2) velocity dispersion, in km\,s$^{-1}$; (3) crossing time; and (4) virial mass, in log\,(${M_\odot}$).}
    \label{tab:dyn_tab}
\end{table}

\subsection{Dynamics and environment as a function of galaxy morphology}
\label{Sec:res-dyn-morpho}

The dynamical parameters provide a quantification of the evolutionary stage of the triplets, based on the configuration of the systems in terms of relative apparent positions and velocities between the member galaxies. Moreover, the environmental parameters quantify the local and LSS environments of the systems. We can explore therefore if there is any relation between the configuration of the SIT triplets and their environments. Since the evolutionary processes of the galaxies in the system leave an imprint in the morphology of its members, we also explore these relations as a function of the morphology of the triplets, based on the four categories defined in Sect.~\ref{Sec:morf}. 
We did not find any relation between the dynamical parameters ($R_H$, $\sigma_{v_r}$, $H_0 t_c$, and $M_{vir}$) and the environment parameters project number density ($\eta_{k,LSS}$), tidal strength of LSS ($Q_{LSS}$), or even distance to nearest neighbour $d_{nn}$. We did not find either a correlation between $\sigma_{v_r}$ and $M_{vir}$ with the $Q_{trip}$ (neither $Q_{A,trip}$). To avoid distraction we do not show the corresponding scatter plots.

\begin{figure}
    \centering
    \includegraphics[width=\columnwidth]{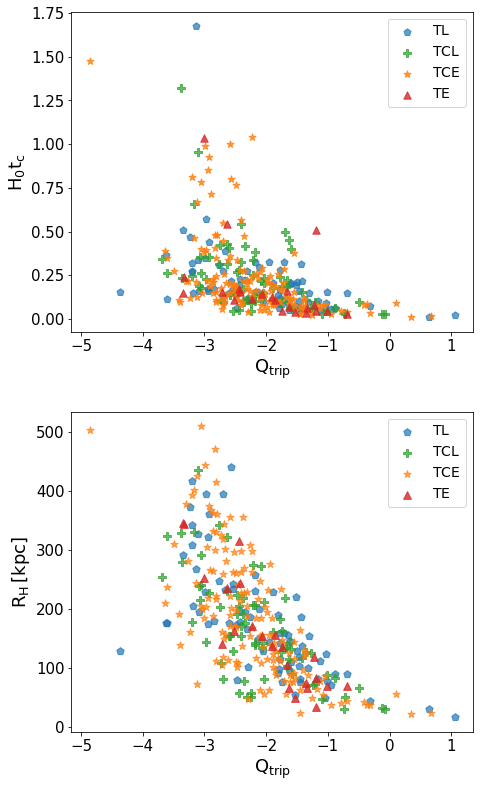}
    \caption{Dependence of the dynamical parameters $H_0 t_c$ and $R_H$ with respect to the local environment. The morphology for each triplet is represented as in Fig.~\ref{fig:LSS}. \textit{Upper panel}: Adimensional crossing time dynamic parameter $H_0 t_c$ with respect to the total local tidal strength $Q_{trip}$. \textit{Lower panel}: Harmonic radius $R_H$ dynamic parameter, in kpc, with respect to the total local tidal strength $Q_{trip}$.}
    \label{fig:Dyn-Qtrip-morpho}
\end{figure}

However we did observe a trend for $R_H$ and $H_0 t_c$ with respect to the local tidal strength $Q_{trip}$, as well as with the $Q_{A,trip}$ parameter. We show this relation in the two panels of  Fig.~\ref{fig:Dyn-Qtrip-morpho}. In general $R_H$ decreases with higher value of the local tidal strength, however systems with the lowest $R_H$ spawn a large range of tidal strength values. This relation is even more present when considering the $H_0 t_c$ versus the tidal strength, where larger values of the $H_0 t_c$ are only present in system with smaller tidal strength. We also found a relation with the morphology of the triplet. We discuss these results in Sect.~\ref{Sec:dis-dyn-morpho}.

\subsection{SFR and mass as a function of galaxy morphology}
\label{Sec:res-sfr-morpho}

Another method to explore the evolution of galaxies is through their SFR. To complement the information we can obtain for the SIT systems regarding their dynamical configuration, we used the SFR-stellar mass ($M_\star$) diagram, which provides information of the evolutionary stage of a system in terms of its global star-formation. 

\begin{figure}
    \centering
    \includegraphics[width=\columnwidth]{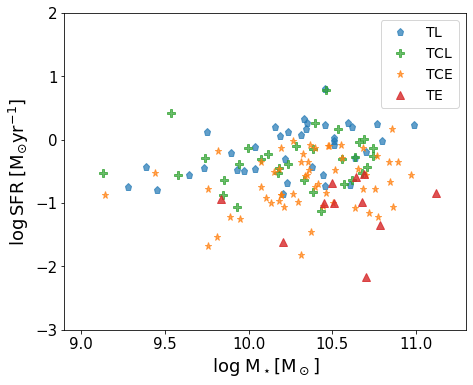}
    \caption{Global SFR-M$_{\star}$ diagram for SIT triplets. The points correspond to galaxy triplets, where the morphology for each system is represented as in Fig.~\ref{fig:LSS}. Star formation rate (SFR) units are on solar masses per year ($\rm M_{\odot} yr^{-1}$) and stellar mass ($M_{\star}$) on solar masses ($M_{\odot}$). We do not show five galaxies with very low SFR.} 
    \label{fig:SFR-M-trip}
\end{figure}

\begin{figure}
    \centering
    \includegraphics[width=\columnwidth]{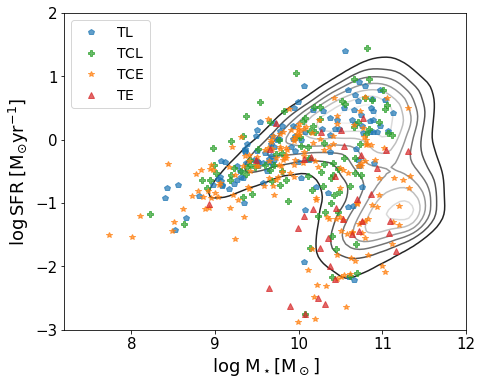}
    \caption{SFR-M$_{\star}$ diagram for SIT galaxies. Marker colour and style indicate the morphology of the triplet where the galaxy corresponds, as in Fig.~\ref{fig:LSS}. Black contours correspond to galaxies in the GSWLC-M2 catalogue \citep{2018ApJ...859...11S}. Star formation rate (SFR) units are on solar masses per year ($\rm M_{\odot} yr^{-1}$) and stellar mass ($M_{\star}$) on solar masses ($M_{\odot}$). Discarding five points with very low SFR}
    \label{fig:SFR-M-A}
\end{figure}

As previously mentioned in Sect.~\ref{Sec:sfr}, we have SFR information for a subsample of SIT triples, based on the GSWLC-M2 catalogue (38 TL triplets, 13 TE triplets, 33 TCL triplets, and 58 TCE triplets). For those triplets, we followed the work of \citet{2013MNRAS.433.3547D} to compute the global SFR-$M_\star$ diagram for galaxy triplets in the SDSS. We therefore compute the mean stellar mass content (defined as $M_{\star, trip}~=~(M_{\star,A}~+~M_{\star, B}~+~M_{\star, C}$)\,$\times$\,3$^{-1}$) and the mean star formation activity (defined as SFR$_{trip}$~=~(SFR$_{A}$~+~SFR$_{B}$~+~SFR$_{C}$)\,$\times$\,3$^{-1}$). The global SFR-$M_\star$ diagram for SIT triplets is shown in Fig.~\ref{fig:SFR-M-trip}. For comparison, the SFR-$M_\star$ diagram for SIT galaxies with respect to galaxies in the GSWLC-M2 catalogue is shown in Fig.~\ref{fig:SFR-M-A}. We found a stratified relation depending on the morphology of the triplets, where TL triplets show the highest global SFR at a given stellar mass, as expected, followed by TCL and TCE triplets, respectively, with TE triples showing the lowest values of the global SFR at a given  stellar mass. We further discuss these results in Sect.~\ref{Sec:dis-sfr-morpho}.

\section{Discussion}
\label{Sec:dis}

\subsection{Environment of galaxy triplets}
\label{Sec:dis-environment}

The introduction of the new parameter $Q_{trip}$ provides a more complete information of the tidal strengths within the system. It considers not only the local tidal strengths exerted on the A galaxy ($Q_{A,trip}$), as defined in \citet{2015A&A...578A.110A}, but also the local tidal strengths exerted on the B ($Q_{B,trip}$) and the C ($Q_{C,trip}$) galaxies, as defined in Eq.~\ref{Eq:Qtrip}. This definition allows us to interpret any correlation of the dynamical parameters of the triplets with the local environment. 

According to Fig.~\ref{fig:Qtrip}, in general, there is a good agreement with the classic $Q_{A,trip}$ parameter (correlation coefficient: 0.71). However, at higher tidal strengths the $Q_{A,trip}$ values are higher than the $Q_{trip}$ (as shown in the lower panel of Fig.~\ref{fig:Qtrip}). This difference is mainly due to closer projected distances of the B and/or C galaxies to the A galaxy, which affect more the increment of the value of the $Q_{A,trip}$ parameter than higher stellar mass ratio, as defined in Eq.~\ref{Eq:Q}. This effect is smoothed when considering $Q_{trip}$. In addition, the distribution of $Q_{trip}$ is narrower ($\sigma_{Q_{trip}}$~=~0.83) than $Q_{A,trip}$ ($\sigma_{Q_{A,trip}}$~=~1.10), reducing the number of outliers (as shown in the upper panel of Fig.~\ref{fig:Qtrip}).

We have observed that there is no dependency on the $Q_{trip}$ with the morphology of the triplet, however we noted that triplets with higher $Q_{trip}$ present signs of interaction that can be detected in the optical images from a visual inspection. Relaxed triplets, without strong interactions between their member galaxies, present lower values of $Q_{trip}$, with higher values when stronger and visible interactions are happening in the system, as can be seen for instance in SIT~190 (see the right lower panel of Fig.~\ref{fig:mergers}). This relation is not very clear when using $Q_{A,trip}$ alone. We therefore conclude that, even if $Q_{A,trip}$ provides a valuable general quantification of the local environment of the triplets, the $Q_{trip}$ parameter, considering the total local tidal strengths for each galaxy in the system, allows a more appropriate exploration of the effects environment on physical properties of the systems. 

In general, we also do not find a relation on the LSS environment with the morphology of the galaxies in the triplets. However, as shown in the distribution of the $Q_{\rm LSS}$ in Fig.~\ref{fig:LSS}, TCE triplets present lower values of this parameter while presenting comparable values of the $\eta_{k,LSS}$. This might indicate that TCE triplets are surrounded by low mass neighbour galaxies in their LSS environment. 

\subsection{Dynamics of galaxy triplets}
\label{Sec:dis-dynparam}

We investigated if there is any connection between the morphology of the galaxies in isolated galaxy triplets and the dynamical evolution of the systems. According to \citet{1982ApJ...255..382H}, compact groups mainly composed of elliptical galaxies are more compact, with lower values of harmonic radius ($R_H$) than in compact groups with more diverse morphologies, and have lower values of crossing time ($H_0 t_c$). Our results are in agreement with \citet{1982ApJ...255..382H} in the sense that TE triplets have the smallest $R_H$ values, indicating that isolated triplets composed of three elliptical galaxies are in general more compact than triplets with one or more spiral galaxies. TE triplets correspond to about 9\% of the sample. For comparison, \citet{2019MNRAS.482.2627T} found that only the 3\% of isolated triplets can be considered as compact systems, however they do not distinguish between galaxy morphology and their criterion to select compact system is much more restrictive. Regarding the $H_0 t_c$ parameter, the median value for TE triplets is not very different to the values for the other morphological categories, however the interquartile range points towards lower values of $H_0 t_c$, as shown in the lower left panel of Fig.~\ref{fig:violin}. 

We found that isolated triplets with a central elliptical galaxy (TCE triplets), which are predominant in the SIT (47.0\%), show higher median values of $R_H$, $\sigma_{v_r}$, and $M_{vir}$ than any other morphological category (see Tab.~\ref{tab:dyn_tab}). This might be due to the fact that elliptical central galaxies are usually more massive than central spirals, therefore physically bound satellite galaxies can be found at larger projected distances, or higher velocity dispersion, as expected by their escape velocity \citep{2014A&A...564A..94A}. Neighbour galaxies with higher velocity may evade the gravitational attraction of the A galaxy and not being captured within the system. Note that we do not observe this trend in TE triplets because galaxies inside these systems might have been evolving cooperatively for a longer time, supported by the previous findings. Our results suggest that triplets composed of three early-type galaxies might be already relaxed systems with dynamical properties slightly different than in triplets presenting late-type galaxies. 

Considering the previous findings for TCE triplets with respect to their LSS environment, we propose the scenario where TCE isolated triplets are primarily formed when an early-type galaxy is located in an environment mainly composed of low mass neighbour galaxies. These neighbours might be physically bound with the early-type galaxy when they are captured by its gravitational potential

\subsection{Dynamics and environment as a function of galaxy morphology}
\label{Sec:dis-dyn-morpho}

In general, we did not find any dependency of the dynamical parameters with the LSS environment, in agreement with \citet{2019MNRAS.482.2627T}. We neither found any clear relation between the local environment and the velocity dispersion or the virial mass of the triplet. This result is interesting since by definition, $Q_{\rm trip}$ strongly depends on the stellar mass. However we have found a dependency of $Q_{\rm trip}$ with $R_H$, and by consequence with $H_0t_c$ (see the upper and lower panels in Fig.~\ref{fig:Dyn-Qtrip-morpho}). Isolated triplets with higher values of $R_H$ and $H_0t_c$ usually have $Q_{\rm trip}<-2$, therefore the tidal strength by the triplet members is not strong. This indicates that less compact and younger triplets (i.e. less long time evolved) present low level of local interaction between its members, independently of the morphology of the galaxies.

On the other hand, the triplets with higher $Q_{\rm trip}$ show lower values of $R_H$ and $H_0t_c$, however note that not all triplets with low $R_H$ and $H_0t_c$ have a high $Q_{\rm trip}$. This results suggests that there are several possible configurations for isolated triplets among the ones with lower $R_H$ and $H_0t_c$, i.e. the most compact and evolved triplets according to \citet{1992ApJ...399..353H}, who classified compact groups in strongly interacting systems and less interacting systems. 

We found that TE triplets generally present low values of $R_H$ and $H_0t_c$, which indicates that are compact and long time evolved systems, in agreement with \citet{1992ApJ...399..353H}. However, we also found a wider range of morphologies among the systems with low values of $R_H$ and $H_0t_c$. Some TL, TCL, and TCE triplets also present low values of $R_H$ and $H_0t_c$ for all values of $Q_{\rm trip}$ as shown in Fig.~\ref{fig:Dyn-Qtrip-morpho}. In addition, as mentioned in Sect.~\ref{Sec:dis-environment}, galaxies in isolated triplets with higher $Q_{trip}$ values present signs of interaction, such as mergers or tidal tails, that are appreciable in their optical SDSS images. Hence these are not relaxed systems, even  presenting low $H_0t_c$ values. 

\begin{figure}
    \centering
    \includegraphics[width=\columnwidth]{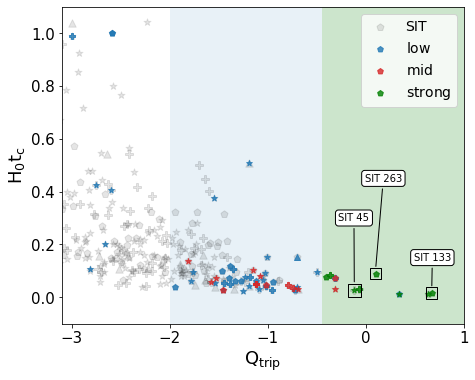}
    \caption{Crossing time vs tidal strength, where we zoomed on the area from $Q_{trip}>-3$ for better visualization. Marker style indicate the morphology of the triplet where the galaxy corresponds, as in Fig.~\ref{fig:LSS}. Gray dots represent the full sample of triplets while blue dots represent galaxies showing signs of interaction, green dots represent galaxies showing slight signs of interaction, and red dots represent galaxies with strong signs of interaction.
Blue colored area represents systems that may have a level of interaction of all types from $-2<Q_{trip}<-0.45$, while the green colored area represents mergers or on-going merger types from  $Q_{trip}>-0.45$.
We specify three example cases that are SIT 263, SIT 45, SIT 133 with a black square. }
    \label{fig:Qtrip_mergerstage}
\end{figure}

We therefore propose to use the $Q_{trip}$ parameter to provide an empirical limit value that help us to identify systems with on-going interactions between their member galaxies. To do so, we performed a visual inspection of the 315 SIT triplets and classify the systems in three categories according to the observed level of interaction of their member galaxies. We classify as 'low interaction' when, without being significantly close to its companion galaxy, there are visible interactions in the galaxy, such as tidal bridges and deformations, which could be due to past interactions in the system, or when there are close pairs with minimal interaction (41 isolated triplets), 'mid interaction' in the case of presence of tidal tails and on-going mergers (19 isolated triplets), and 'strong interaction' when the galaxies are close to the last merging stage of core fusion (9 isolated triplets). We refer to these systems as low, mid, and strong, respectively, in Fig.~\ref{fig:Qtrip_mergerstage}. As shown in this figure, for values of $Q_{trip}~>~-2$ and $H_0t_c~<~1$ we start to observe interactions in some of the systems, mainly low and mid interactions; while SIT triplets with $Q_{trip}>-0.45$ present on-going mergers (mid interactions) or strong interactions.  

\begin{figure*}
\centering 
    \includegraphics[width=0.45\textwidth]{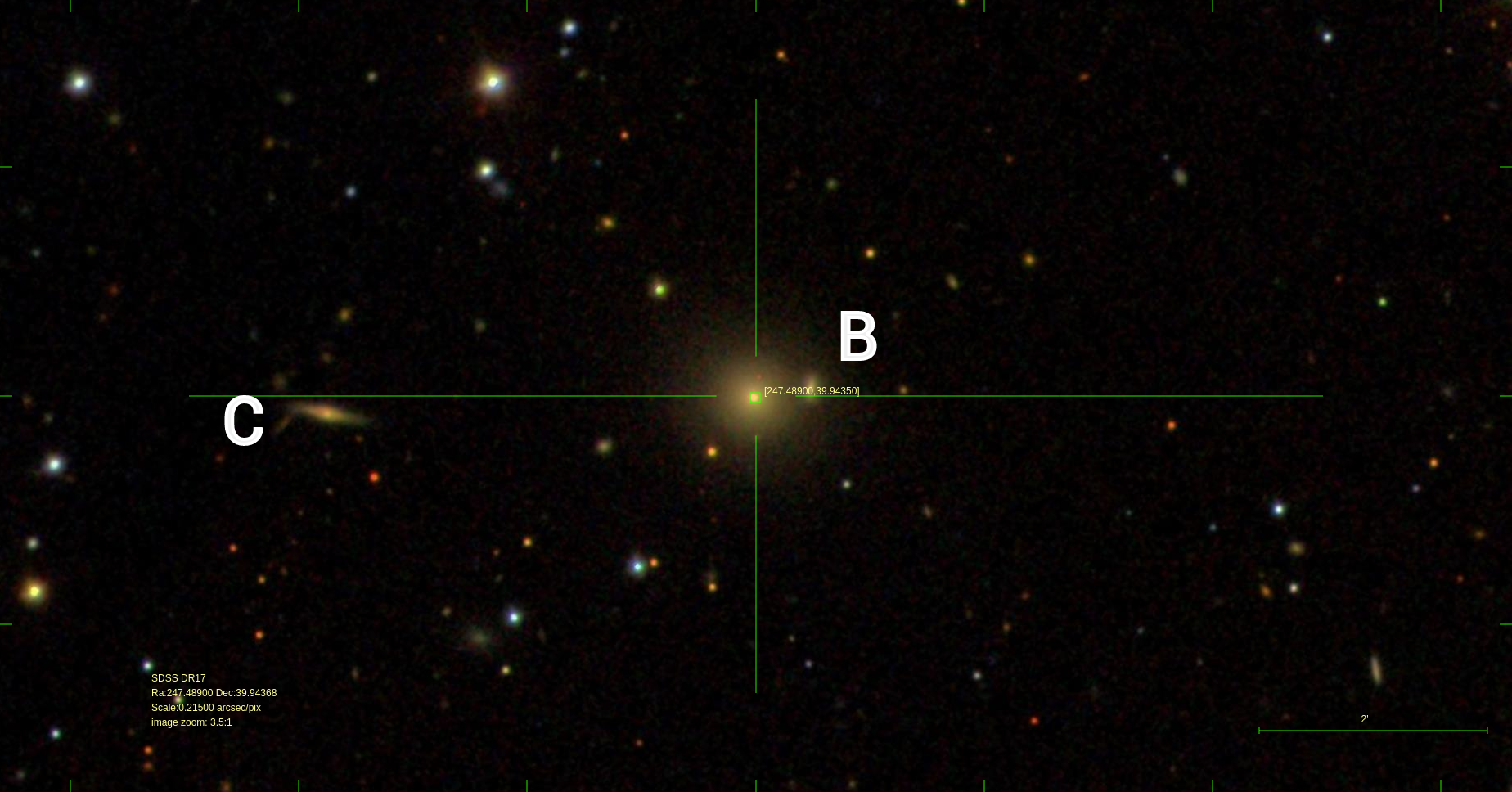}
    \includegraphics[width=0.45\textwidth]{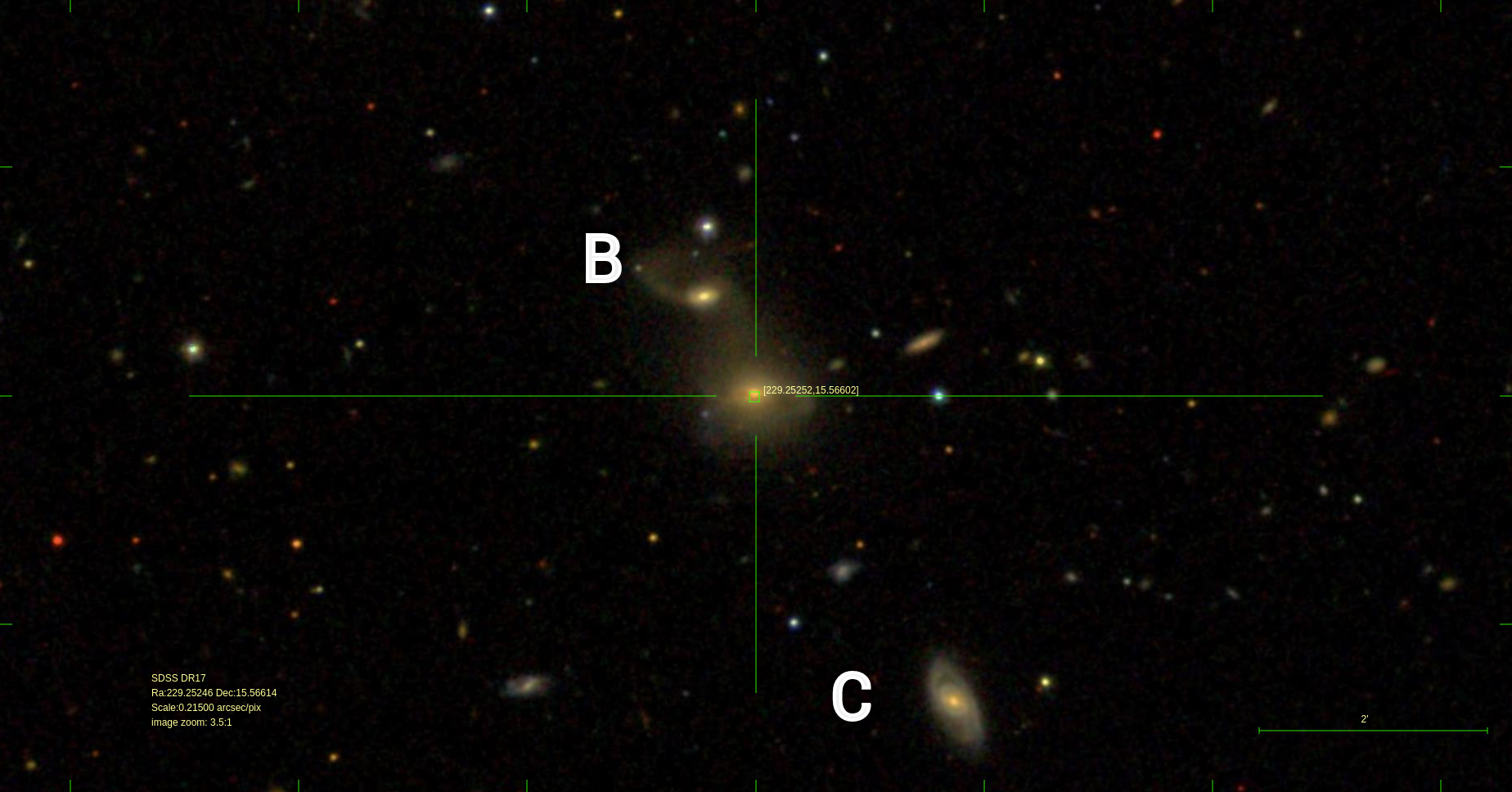} \\
    \includegraphics[width=0.45\textwidth]{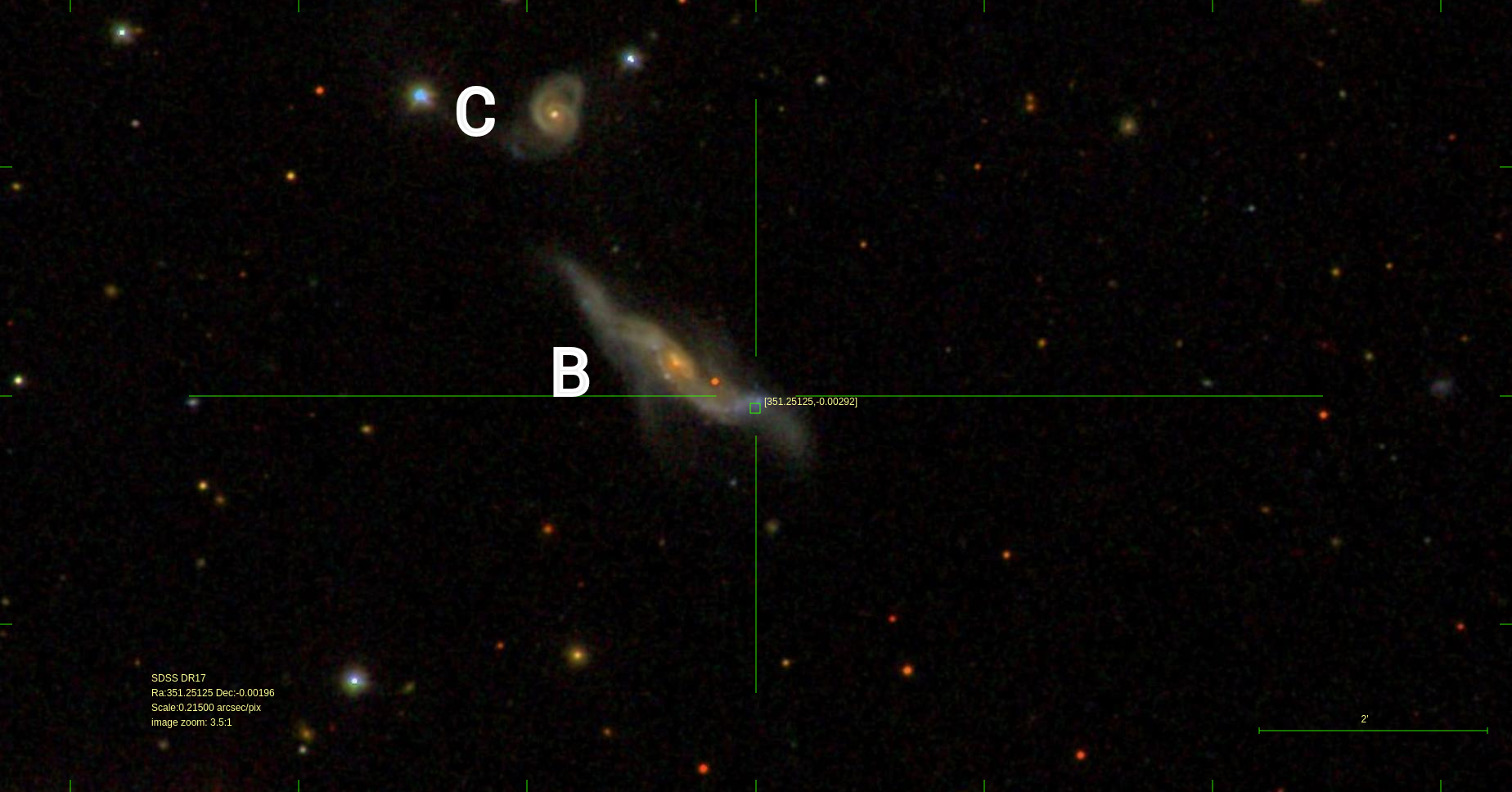}
    \includegraphics[width=0.45\textwidth]{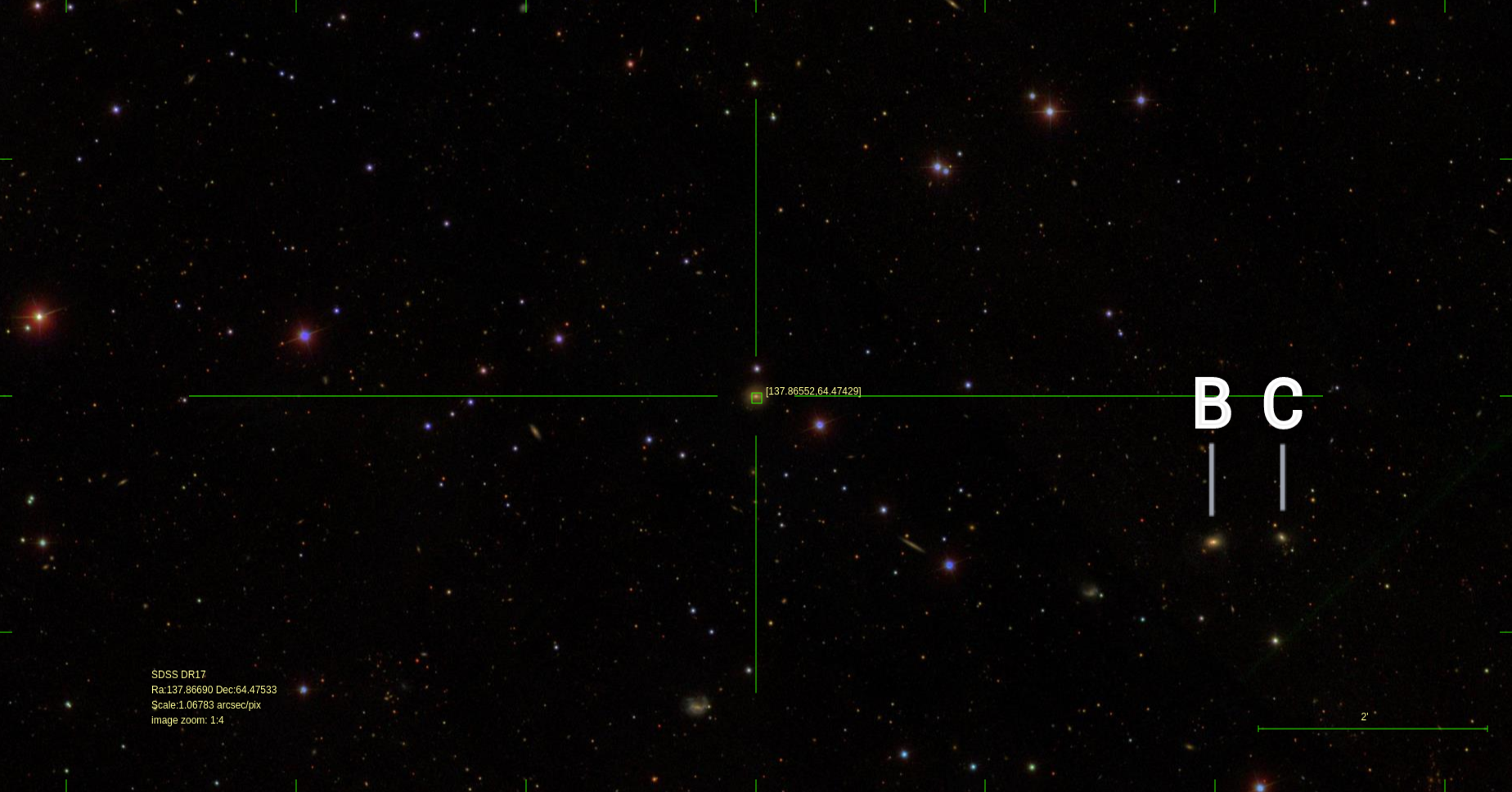}
    \caption[Pair]
    {\small SDSS-DR17 three colour images of three selected SIT triplets as an example of systems with interacting galaxies (image scale 0.215\arcsec/pixel), and one selected triplet as an example of a relaxed galaxy triplet (image scale 1.067\arcsec/pixel). Upper left panel: SIT 133 (TCE triplet). Upper right panel: SIT 263 (TCE triplet). Lower left panel: SIT 45 (TL triplet). Lower right panel: SIT 190 (TE triplet), a relaxed isolated triplet with $Q_{trip}~=~-2.72$ and $H_0t_c~=~0.16$. Since the scale is larger to cover the full triplet, a line has been added for an easier identification of the B and C galaxies and to avoid confusion with background/foreground galaxies.}
    \label{fig:mergers}
\end{figure*}

We have found that, in general, interactions in galaxy triplets occur between two member galaxies, primarily between the A galaxy and another galaxy. An example of isolated triplet in each category is presented in Fig.~\ref{fig:mergers}. There is only one isolated triplet where the three galaxies are interacting, SIT\,45. In a separated work, we consider two scenarios for the present configuration of the triplet, one where the A galaxy in this triplet is a tidal galaxy formed from a previous interaction between the B and C galaxies, and another where the A galaxy arrives to the system after the interaction \citep{2022arXiv220912850G}.

We propose to use the $Q_{trip}$ parameter to classify the dynamical state of isolated triplets with low values of $R_H$ (i.e. the most compact) and $H_0t_c$, since not all these triplets would be long time evolved or relaxed systems. SIT triplets with $Q_{trip}~<~-2$ are relaxed systems, more dynamically evolved, while triplets with $Q_{trip}~>~-2$ show compact configurations due to interactions happening in the systems, such as on-going mergers. This empirical value is in agreement with numerical simulations. According to previous works, the evolution of a galaxy may be affected by external influence when the corresponding tidal force amounts to 1\% of the internal binding force \citep{1984PhR...114..321A,1992AJ....103.1089B}. This theoretical value corresponds to a tidal strength of $Q~=~-2$, which allows to separate the interactions that might affect the evolution of isolated galaxies \citep{2007A&A...472..121V,2013A&A...560A...9A}. 
Note that this empirical value is limited to the magnitude limit of SDSS optical images, where the 95\% completeness limits for each band are ($u$, $g$, $r$, $i$, $z$)~=~(22.15, 22.2, 22.2, 21.3, 20.5), respectively \citep{2000AJ....120.1579Y,2006AJ....131.2332G}. 

\subsection{SFR and mass as a function of galaxy morphology}
\label{Sec:dis-sfr-morpho}

The SFR-$M_\star$ diagram for SIT galaxies in Fig.~\ref{fig:SFR-M-A} shows that most of the galaxies are star-forming galaxies, located along the main sequence. A small number are quiescent galaxies (low SFR-$M_\star$), most of them early-type (since they belong to TE triplets), and galaxies in the green valley below the main sequence. In general, most of the late-type galaxies belonging to TL triplets are star-forming galaxies, but we also find a few of these galaxies (four galaxies) in the green valley. Similarly, about a dozen of early-type galaxies in TE triplets are located in the green valley and the main sequence. This result is expected since \citet{2022arXiv220913437P} found that 47\% of early-type galaxies in their sample present ongoing star-formation, with SFRs comparable to star-forming late-type galaxies. They also found that, in general, star-forming galaxies are mainly found in low-density environment, therefore it is expected to find star-forming early-type galaxies in isolated triplets. 

As mentioned in Sect.~\ref{Sec:res-sfr-morpho}, we followed the work developed by \citet{2013MNRAS.433.3547D} to derive the global SFR-$M_\star$ diagram for SIT triplets presented in Fig.~\ref{fig:SFR-M-trip}. The upper panel of Fig.~6 in \citet{2013MNRAS.433.3547D} shows a stratified relation where blue triplets show the highest global SFR at a given stellar mass, while red triplets show the lowest values of the global SFR at a given  stellar mass. We have confirmed that this stratification also exists when considering the morphology of the triplets. TL triplets show the highest global SFR at a given stellar mass, followed by TCL and TCE triplets, respectively, with TE triples showing the lowest values. We have checked that this stratification disappears when considering only the SFR and M$_\star$ for the A galaxy in the triplet. This might indicate that the physical properties of galaxy triplets, such as colour or SFR, do not depend on the properties of the central galaxy, but on the global properties considering the three galaxies in the system. Therefore, there would be no dominant galaxy in triplets in terms of properties of stellar populations such as colour and SFR. It this sense, it would be interesting to investigate if similar results are found when considering other properties, for instance nuclear activity.

\begin{figure}
    \centering
s    \includegraphics[width=\columnwidth]{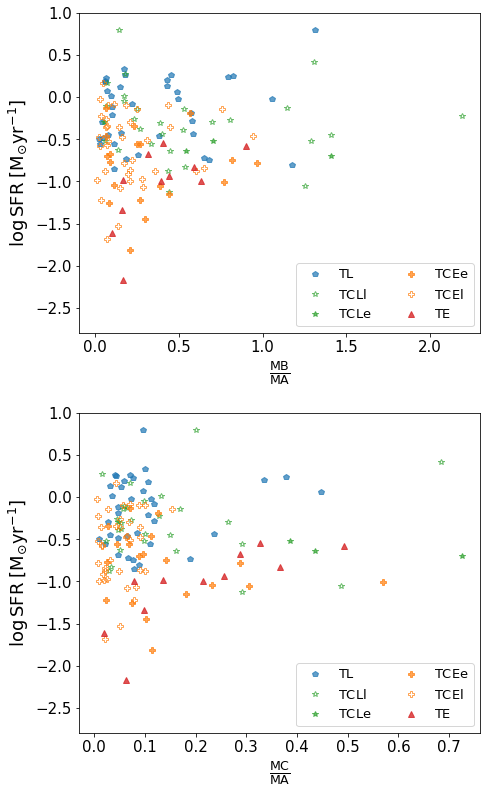}
    \caption{Global SFR for SIT triplet versus stellar mass ratio. The upper panel shows the global SFR versus the mass ratio of the B galaxy with respect to the A galaxy ($M_{\star,B}$/$M_{\star,A}$) and the lower panel shows the global SFR versus the mass ratio of the C galaxy with respect to the A galaxy ($M_{\star,C}$/$M_{\star,A}$). The morphological categories represented in the legend are as follows: blue diamonds correspond to triplets with three late-type galaxies (TL), green unfilled stars correspond to triplets with two late-type galaxies (with the central late-type galaxy) and one early-type (TCLl), green stars to triplets with the central late-type galaxy and two early-type galaxies (TCLe), unfilled orange crosses represent triplets with two late-type galaxies and the central early-type (TCEl), orange diamonds correspond to triplets with two early-type galaxies (with the central early-type galaxy) and one late-type (TCEe), and triplets with three elliptical galaxies (TE) are depicted by red triangles.}
    \label{fig:SFR-Mb-Mc}
\end{figure}

Since C and B galaxies are usually less massive than A galaxy \citep{2015A&A...578A.110A}, we further explored the observed stratification relation of global SFR with morphology of the triplet in terms of the stellar mass ratio of the B and C galaxies with respect to the A galaxy. As shown in Fig.~\ref{fig:SFR-Mb-Mc}, this stratification is still present. Moreover, we find that the global SFR in TE triplets increases with mass ratio. To better understand this trend, new more refined morphological subcategories were defined for TCL and TCE triplets, taking into account the number of galaxies with a certain morphology. If TCL and TCE triplets have two late-type galaxies and one early-type galaxy, an 'l' was added to the name, while if the triplets contain two early-type galaxies and one early-type galaxy, an 'e' was added, resulting in the following categories:
\begin{itemize}
    \item TL: Triplet with three late-type galaxies, 38 triplets.
    \item TCLe: Triplet with the central galaxy as late-type, while their companions are a late-type and early-type galaxies, 5 triplets.
    \item TCLl: Triplet with the central galaxy as late-type, while their companions are both late-type galaxies, 28 triplets.
    \item TCEe: Triplet with the central galaxy as early-type, while their companions are a late-type and early-type galaxies, 20 triplets.
    \item TCEl: Triplet with the central galaxy as early-type, while their companions are both late-type galaxies, 38 triplets.
    \item TE: Triplet with three early-type galaxies, 13 triplets.
\end{itemize}

We found that this trend is also observed in TCEe triplets. The more similar the stellar mass of the B and C galaxy is with respect to the stellar mass of the A galaxy, the higher the global SFR in TE and TCEe isolated triplets. For the rest of the morphological categories we do not observe any relation. 

\citet{2010MNRAS.404.1775T} suggests that elliptical galaxies in low density environments may be rejuvenated by eventually accreting cold gas. This rejuvenation phase might be mainly due to galaxy mergers and interactions, with a subsequent transition into a secular evolution phase \citep{2004ARA&A..42..603K}. In fact, a number of the star-forming early-type galaxies in \citet{Paspaliaris2022} show signs of interactions.

Our results suggest that, when isolated triplets are composed of two or more early-type galaxies, the global SFR of the system increases with the stellar mass ratio of the galaxies with respect to the central A galaxy, therefore the system is globally 'rejuvenated'. This might be triggered by the gas supply being accreted from the companion galaxy.

\section{Summary and conclusions} 
\label{Sec:con}

We present a statistical study of the dynamical properties of isolated galaxy triplets as a function of their environment, exploring the connections of the dynamical evolution of the systems, with respect to the evolution of the galaxies composing the triplets. We therefore consider observational properties such as morphology and star formation rate (SFR).

We used the SDSS-based catalogue of Isolated Triplets  \citep[SIT,][]{2015A&A...578A.110A} which contains 315 isolated galaxy triplets. We classified each triplet according to galaxy morphologies using the classification performed by \citet{2018MNRAS.476.3661D}. SIT triplets were classified as TL, where the three galaxies are late-type; TCL, when only the central galaxy in triplets is a late-type galaxy; TCE, when only the central galaxy is an early-type galaxy; and TE when the three galaxies composing the triplet are early-type galaxies. 

To quantify the local and LSS environments we used the $Q_A$ parameter defined in \citet{2015A&A...578A.110A}, as explained in Sect.~\ref{Sec:env}, and we define a new $Q_{\rm trip}$, which quantifies the total local tidal strength in the triplets, as presented in Sect.~\ref{Sec:res-environment}.
To consider the dynamical stage of the system we used the parameters harmonic radius ($R_H$), velocity dispersion ($\sigma_{vr}$), crossing time ($H_0t_c$), and virial mass ($M_{vir}$) as explained in Sect.~\ref{Sec:dyn}. 

In general, we did not find any dependency of the dynamical parameters with the LSS environment. We neither found any clear relation between the local environment and the velocity dispersion or the virial mass of the triplet. However we have found a dependency of $Q_{\rm trip}$ with $R_H$, and by consequence with $H_0t_c$. For a complete analysis, we investigated if star formation in the system is affected by its morphology, environment, and dynamical stage. 
We found a stratified relation depending of the morphology of the triplets, where TL triplets show the highest global SFR at a given stellar mass, followed by TCL and TCE triplets, respectively, with TE triples showing the lowest values of the global SFR at a given  stellar mass.

We found the following main conclusions: 

\begin{enumerate}

    \item The local tidal strength environment parameter on central galaxy $Q_{\rm A,trip}$ provides a valuable general quantification of the local environment of the triplets, however, the $Q_{\rm trip}$ parameter considering the total local tidal strengths for each galaxy in the system, allows a more appropriate exploration of the effects environment on physical properties of the systems.
    
    \item In general, we do not find a relation in the local and in the LSS environment with respect to the morphology of the triplets. However, we found that TCE triplets might be mainly surrounded by low mass neighbour galaxies in their LSS environment. 
    
    \item We found that TCE triplets, show higher median values of $R_H$, $\sigma_{v_r}$, and $M_{vir}$ than any other morphological category. This might be due to the fact that elliptical central galaxies are usually more massive than central spirals, therefore physically bound satellite galaxies can be found at larger projected distances, or higher velocity dispersion, as expected by their escape velocity.
    
    \item TE triplets are in general more compact than triplets with one or more late-type galaxies. Their dynamical parameters also indicates that these are relaxed systems, where their galaxies might have been evolving cooperatively for a long time.
    
    \item The LSS environment does not have an influence in the dynamical configuration of the isolated triplets. However, there is a dependence with the local environment. Less compact and younger triplets (i.e. less long time evolved) present low level of interaction between its members, independently of the morphology of the galaxies. 
    
    \item We can use the $Q_{trip}$ parameter to classify the dynamical state of isolated triplets with low values of $R_H$ and $H_0t_c$. SIT triplets with $Q_{trip}~<~-2$ are relaxed systems, more dynamically evolved, while triplets with $Q_{trip}~>~-2$ show compact configurations due to interactions happening in the systems, such as on-going mergers.
    
    \item Exist a stratification in the global SFR of the triplets, which disappears when considering only the properties of the central galaxy. Therefore, there is no dominant galaxy in triplets in terms of properties of stellar populations such as colour and SFR.
    
    \item We found that the global SFR in isolated triplets composed of two or more early-type galaxies increases with the stellar mass ratio of the galaxies with respect to the central A galaxy, therefore the system is globally 'rejuvenated'.
    
\end{enumerate}

Considering our findings for TCE triplets with respect to their LSS environment and dynamical parameters, we propose the scenario where TCE isolated triplets are primarily formed when an early-type galaxy is located in an environment mainly composed of low mass neighbour galaxies. These neighbours might physically be bound with the early-type galaxy if are captured by its gravitational potential. 
This would be the main formation mechanism in isolated triplets, since TCE triplets are the predominant morphology in the SIT (47\%), with a percentage even larger than expected if galaxies are distributed randomly in terms of their morphology (38\% early-type and 62\% late-type).

\begin{acknowledgements}
We thank the anonymous referee for the constructive revision of this work.

PVB and MAF acknowledge financial support by the DI-PUCV research project 039.481/2020. MAF also acknowledges support from FONDECYT iniciaci\'on project 11200107 and the Emergia program (EMERGIA20\_38888) from Consejería de Transformación Económica, Industria, Conocimiento y Universidades and University of Granada. 
SV acknowledges support from project P20\_00334 financed by the Junta de Andaluc\'ia and from FEDER/Junta de Andaluc\'ia-Consejer\'ia de Transformaci\'on Econ\'omica, Industria, Conocimiento y Universidades/Proyecto A-FQM-510-UGR20. 
SDP is grateful to the Fonds de Recherche du Québec - Nature et Technologies and acknowledges financial support from the Spanish Ministerio de Econom\'ia y Competitividad under grants AYA2016-79724-C4-4-P and PID2019-107408GB-C44, from Junta de Andaluc\'ia Excellence Project P18-FR-2664, and also acknowledges support from the State Agency for Research of the Spanish MCIU through the `Center of Excellence Severo Ochoa' award for the Instituto de Astrof\'isica de Andaluc\'ia (SEV-2017-0709).

This research made use of \textsc{astropy}, a community-developed core \textsc{python} ({\tt http://www.python.org}) package for Astronomy \citep{2013A&A...558A..33A}; \textsc{ipython} \citep{PER-GRA:2007}; \textsc{matplotlib} \citep{Hunter:2007}; \textsc{numpy} \citep{:/content/aip/journal/cise/13/2/10.1109/MCSE.2011.37}; \textsc{scipy} \citep{citescipy}; and \textsc{topcat} \citep{2005ASPC..347...29T}. This research made use of \textsc{astrodendro}, a Python package to compute dendrograms of Astronomical data (http://www.dendrograms.org/).

This research has made use of the NASA/IPAC Extragalactic Database, operated by the Jet Propulsion Laboratory of the California Institute of Technology, un centract with the National Aeronautics and Space Administration.

Funding for SDSS-III has been provided by the Alfred P. Sloan Foundation, the Participating Institutions, the National Science Foundation, and the U.S. Department of Energy Office of Science. The SDSS-III Web site is http://www.sdss3.org/. The SDSS-IV site is http://www/sdss/org. Based on observations made with the NASA Galaxy Evolution Explorer (GALEX). GALEX is operated for NASA by the California Institute of Technology under NASA contract NAS5-98034. This publication makes use of data products from the Wide-field Infrared Survey Explorer, which is a joint project of the University of California, Los Angeles, and the Jet Propulsion Laboratory/California Institute of Technology, funded by the National Aeronautics and Space Administration.
\end{acknowledgements}

%
%
\bibliography{References}
\bibliographystyle{aa}

\end{document}